\documentclass[aps,prd,a4paper,twocolumn,amsmath,showpacs,superscriptaddress,nofootinbib,preprintnumbers]{revtex4-1}

\usepackage{verbatim}
\usepackage[T1]{fontenc}
\usepackage[utf8]{inputenc}
\usepackage[american]{babel}
\usepackage{epsfig}
\usepackage{graphicx}
\usepackage{booktabs}
\usepackage{multirow}
\usepackage{dcolumn}
\usepackage{amsmath}
\usepackage{mathtools}
\usepackage{amsfonts}
\usepackage{amssymb}
\usepackage{epstopdf}
\usepackage{bm}
\usepackage{siunitx}
\usepackage{braket}
\usepackage{enumitem}
\usepackage{soul}
\usepackage[table]{xcolor}
\usepackage{color}
\usepackage{transparent}
\usepackage{pifont}

\def\simpropto{\lower.2ex\hbox{$\; \buildrel \propto \over \sim \;$}}
\def\ltsim{\lower.5ex\hbox{$\; \buildrel < \over \sim \;$}}
\def\gtsim{\lower.5ex\hbox{$\; \buildrel > \over \sim \;$}}

\newcommand{\squeezeup}{\vspace{-4.5mm}}

\definecolor{navyblue}{rgb}{0.0, 0.0, 0.5}
\definecolor{royalblue}{rgb}{0.25, 0.41, 0.88}
\definecolor{cadmiumgreen}{rgb}{0.0, 0.42, 0.24}
\definecolor{blue-violet}{rgb}{0.54, 0.17, 0.89}
\definecolor{darkviolet}{rgb}{0.58, 0.0, 0.83}
\definecolor{orange(colorwheel)}{rgb}{1.0, 0.5, 0.0}

\usepackage{hyperref}
\hypersetup{
    colorlinks=true, 
    linkcolor=royalblue, 
    citecolor=magenta}

\newcommand\ee{\end{equation}}
\newcommand\be{\begin{equation}}
\newcommand\eea{\end{eqnarray}}
\newcommand\bea{\begin{eqnarray}}

\usepackage{booktabs}
\usepackage{multirow}
\usepackage{dcolumn}
\usepackage{colortbl}

\definecolor{magenta(process)}{rgb}{1.0, 0.0, 0.56}

\definecolor{darkspringgreen}{rgb}{0.09, 0.45, 0.27}

\definecolor{royalblue(web)}{rgb}{0.25, 0.41, 0.88}

\begin{document}

\title{The limits of cosmology: role of the Moon
 }

\author{Joseph Silk}
\email{silk@iap.fr}
\affiliation{Institut d'Astrophysique de Paris, UMR7095:CNRS \& UPMC-Sorbonne University, F-75014, Paris, France}
\affiliation{Department of Physics and Astronomy, The Johns Hopkins University Homewood Campus, Baltimore, MD 21218, USA}
\affiliation{BIPAC, Department of Physics, University of Oxford, Keble Road, Oxford OX1 3RH, UK}

\date{\today}

\preprint{}
\begin{abstract}

The lunar surface allows a unique way forward in cosmology, to go beyond current limits. 
The far side provides an unexcelled   radio-quiet environment for probing the dark ages via 21 cm interferometry to seek elusive clues on the nature of the infinitesimal fluctuations that seeded galaxy formation. Far-infrared telescopes in  cold and dark lunar polar craters will probe back to the first months of the Big Bang and study associated spectral distortions in the CMB.  Optical  and IR megatelescopes will image  the first star clusters in the universe and seek biosignatures in the atmospheres of unprecedented numbers of nearby habitable zone exoplanets.
The goals are compelling and a stable lunar platform will enable construction of telescopes that can access trillions of modes in the sky,  providing the key to  exploration of  our cosmic origins.

\end{abstract} 
\maketitle

\section{Introduction}
Cosmological advances are proving difficult to maintain with future experiments. There is no detection of dark matter
\cite{2020arXiv200106193P}, dark energy remains indistinguishable from a cosmological constant \cite{2018RPPh...81a6901H}, and the tensor-to-scalar ratio  in the cosmic microwave background fluctuations as a verification of inflation remains elusive and even unreachable for  more general, non-Planck scale  inflation  models \cite{2019BAAS...51c.338S}.

Perhaps we should also be looking elsewhere for potential progress in cosmology. I develop here the case for building  lunar telescopes of unprecedented size in order to tackle  frontier science in  cosmology and astrophysics. 
Two key questions  that are central to cosmology and astrophysics can be addressed by lunar megatelescopes. These are: where did we come from, and are we alone in the universe? 

The dark ages represent the ultimate frontier that is directly accessible with conventional telescopes. The lunar environment provides a stable platform for the astronomy of the future. Here we can detect the gaseous building blocks that initiated structure formation and probe the 21cm signals for signs of
 primordial non-gaussianity, arguably the only  robust predictor of generic inflation models \cite{2019BAAS...51c.107M}. 
 We can build far infrared spectrometers capable of probing the inevitable energy input associated with fluctuation growth from the first years of the universe
 \cite{2019arXiv190901593C}. We can build 100m class telescopes, and  consider km-aperture infrared interferometers, to provide imaging capability of exoplanets and of the first stars in the universe \cite{2019arXiv190802080S}.
We can  search spectroscopically
 for biological and even technological signatures from many thousands of rocky core, habitable zone exoplanets \cite{2020MNRAS.495.1000S}. We can build gravitational wave interferometers with 100 km baselines that would sample a frequency space band corresponding to a black hole mass regime inaccessible from the Earth or even in space
 \cite{1990AIPC..202..188S}.
 All of these goals can be uniquely advanced with lunar telescopes.

%
%
%

\begin{figure}[!hbt]
%
%

\centering
\begin{center}
\includegraphics[width=0.5\textwidth,
]
{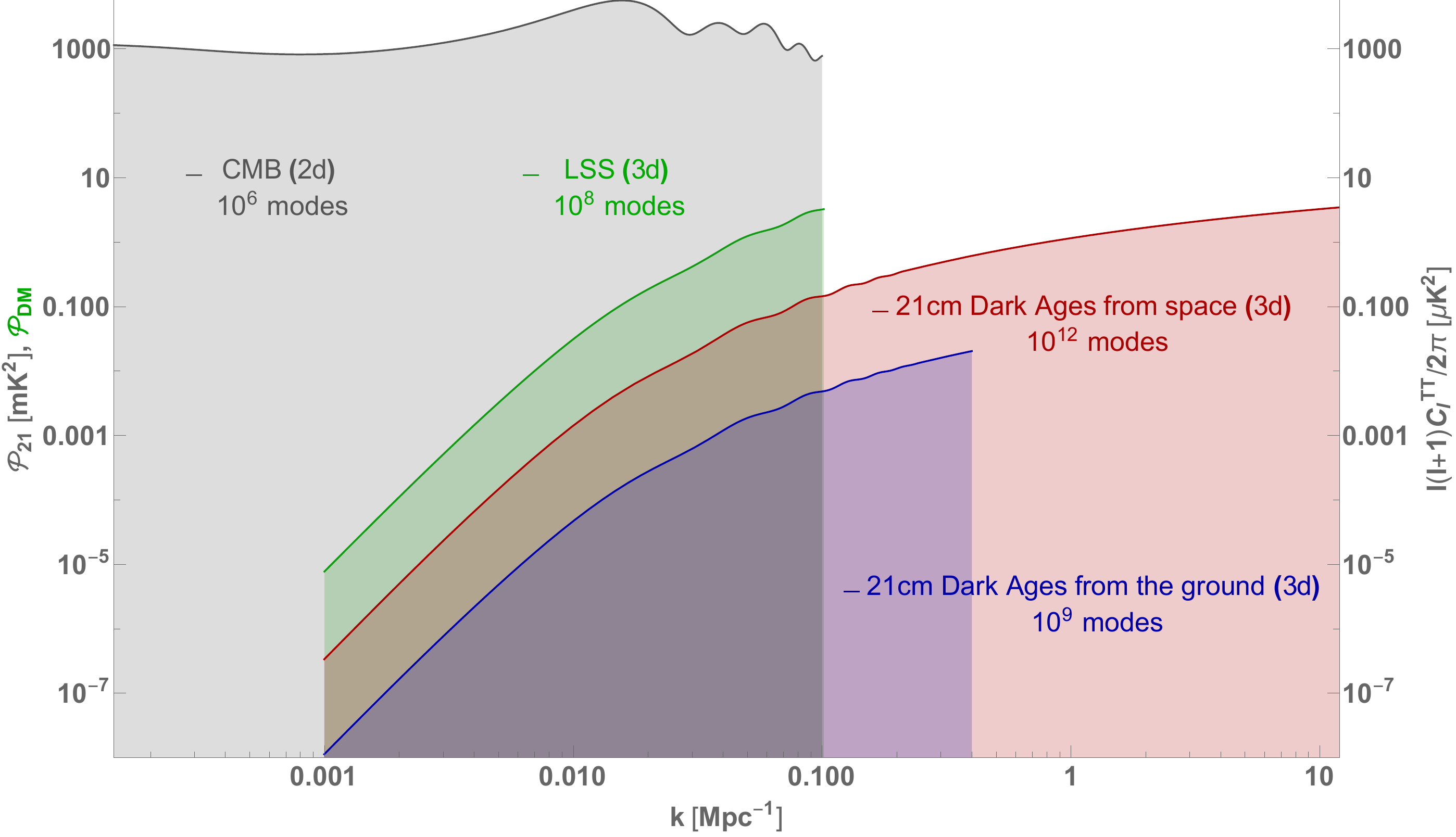}

\caption{
The scope of different cosmological probes for accessing large numbers of
modes \cite{cole}. 
In grey is the TT
angular power spectrum in units of $\mu K^2$ divided by a factor of 1000 to be visually comparable to SDSS and 21 cm. Multipoles are 
roughly mapped to wavenumbers by $\ell =14000k/\rm Mpc.$
In green is the dimensionless 3d matter power spectrum computed with CAMB at redshift 1,
to which large-scale structure probes such as LSST and EUCLID will be sensitive on scales
between $k\sim 0.001$ to 0.1 Mpc$^{-1}$,
up to around redshift 2.5. 
In blue is the 3d 21cm power
spectrum in units of mK$^2$ at redshift 27, which is the highest redshift accessible from ground-based
experiments such as HERA and SKA. In red is the 3d 21cm power spectrum in units of mK$^2$ at
redshift 50, which would be accessible from the far side of the Moon. 
}
\label{figcole1}
\end{center}

%
\end{figure}

 The first challenge to be addressed with lunar telescopes will most likely be that of  
 the last unexplored frontier in astronomy, the dark ages. Pristine clouds of hydrogen are building blocks of the future but also direct witnesses of the past. Very low frequency radio astronomy is uniquely able to tune into their testimony about the beginning of creation. These early clouds have spin temperatures  controlled by atom collisions and are colder than the cosmic microwave background when we observe them after recombination. This is before any stars, galaxies or quasars have formed at very high redshift and generated  Lyman alpha photons whose  hydrogen excitations  would have modified the  spin temperature. 21cm line mapping and tomography at frequencies of 10-50 MHz enables  access to previously inaccessible parameter space, sampling a large number of independent modes describing primordial density fluctuations (Fig.~\ref{figcole1}). 
 
Critical for such  advances is access to small scales $k\sim 10 \rm Mpc^{-1}.$ 
These may be feasible on a lunar site and with adequate frequency resolution,
as described in these proceedings \cite{2020arXiv200713353M}.
A lunar platform promises  to  open a new window on the properties of the early universe and complement both future (e.g. LiteBIRD/CMB-S4) CMB  temperature and  polarization signals, and  large-scale structure studies with  next generation telescopes on the ground  (e.g. SKA/LSST)  and in space (e.g. JWST/EUCLID/RST), as well as open new reaches in habitable exoplanet yields.


\section{Inflation}
The nongaussianity parameter $ f_{NL}$ is robustly predicted  in single field inflation, 
and is generic to  multifield inflation, albeit potentially very small. 
Because of the stringent  Planck limits  in  the $r-n_s$ plane \cite{2018arXiv180706211P}, where $n_s$ is the scalar spectral index, viable inflationary models today mostly have  low 
tensor-to-scalar ratio $r. $ This means that CMB polarization experiments need exquisite sensitivity
at the nanoK level compared to the current $\sim \mu$K limit,   wide sky coverage to  combat cosmic variance, and broad spectral resolution to eliminate foregrounds, in order to qualitatively improve our understanding of inflation. 

 Single field slow roll inflationary potentials which reproduce the measured spectral tilt generally predict $r\gtsim 0.0001$ and motivate future experimental searches.
However   these inflationary models are non-generic, and there is no guaranteed prediction of $r$
\cite{2016ARA&A..54..227K}.
 If inflation occurred  at a energy much below $m_{infl}\sim 10^{15}$ GeV, then any primordial inflation-induced tensor mode is of  strength $r\propto (m_{infl}/m_{pl}). $  The tensor-to-scalar ratio  is likely to be small in many inflationary models
\cite{knox}, with indeed an effective limit of  $r\gtsim 10^{-6}$ being required to avoid confusion with the primordial  inflationary tensor signal induced by   2nd order 
scalar-induced B modes \cite{wands}.

Primordial nongaussianity, reviewed  here \cite{celoria},   provides a robust but challenging  complementary probe of inflation. 
The gauge-invariant newtonian potential $\phi_g$ is gaussian but has  higher order correction terms $\phi=\phi_g + f_{NL}  \phi_g^2$ that define the density  fluctuation
three-point function as a direct measure of local nongaussianity.

Even the simplest single field  inflationary models predict small but non-zero deviations from gaussianity,  while many multi-field inflation models are expected to have $f_{NL}\sim 1 $ and many even predict features in $f_{NL}.$ The errors bars on $f_{NL} $ scale with the inverse of the square root of the number of independent modes.   The limited number of modes in the CMB ($N\sim 10^6$) and in large-scale galaxy surveys ($N\sim 10^8$) strongly motivates exploration of  21 cm dark ages cosmology at z= 25-75 (20-60 MHz),  where up to $\sim 10^{12}$  modes can be explored, at least in principle. Attainment of a robust  limit on (or detection of)  primordial nongaussianity would provide the ultimate probe of generic inflation.

The first quantitative estimate of the minimum value of primordial nongaussianity was given for single field  slow roll inflation \cite{maldacena}, with the result that the  three-point functions are determined completely by the tilt of the spectrum of the two-point functions for the scalar mode, enabling the primordial nongaussianity to be expressed in terms of the spectral tilt as  $f_{NL}=-(5/12)(n_s-1). $  While   $f_{NL}$ may be much larger in 
multifield inflation \cite{panagopoulos},  in single field inflation, however,  
there remains at least a minimal effect   of order  the Maldacena conjecture  for local non-gaussianity \cite{2020arXiv200708877M}. One can thereby test inflation, primordial nongaussianity being a guaranteed signal, and even its quantum origin  may be revealed \cite{green,arkani}.


\section{ CMB spectral distortions}
Following the pioneering observations with COBE/FIRAS in the early 1990s, 
the theoretical foundation of spectral distortions has seen major advances in recent years, highlighting the immense potential of this emerging field to
probe the thermal history of the universe from when it was a few months old until today.  The sky-averaged CMB spectrum is known to be extremely close to a perfect blackbody at a temperature \citep{1996ApJ...473..576F}
$T_0 = 2.7255\pm 0.0006$ K, with possible distortions limited to  parts in $10^5$.

 Spectral distortions from the predicted blackbody spectrum of the cosmic microwave  background, the fossil radiation from the Big Bang, are an important tool for understanding
 the physics of recombination and reionization, the origin of structure, and the origin of  the CMB itself.
The required high precision measurements of the CMB energy spectrum would open a new window into the physics of the early Universe, constraining cosmological models  and fundamental physics in ways not possible using other techniques, most notably by searching for blackbody spectral distortions.

This new horizon for cosmology urgently merits  serious exploration in part as a  consequence of  the uncertain prospects for detecting a primordial B-mode signal from inflation. It is especially 
 relevant to therefore consider complementary approaches that are capable of yielding unique information on the primordial universe,  directly probing unprecedentedly early epochs back to the epoch when the cosmic blackbody radiation originated or even all the way back to the end of inflation. This goal can only come 
 from the diffuse cosmological backgrounds of cosmological neutrinos, gravitational waves or relic photons. Of these,  neutrino detection, while possible in principal,  remains highly  futuristic \citep{2019JCAP...07..047B},
 and measuring spectral distortions 
 would be a vastly more remote goal. Direct detection of a  stochastic gravitational wave background is plagued by instrumental noise \citep{2020MNRAS.491.4690M}, and predicted spectral distortions of the relic photon background radiation  are seen against immense far infrared  foregrounds.
 We can still be optimistic about feasibility, in large part because  we have been here before in terms of overcoming foregrounds.  The unprecedented limits on  infinitesimal deviations from a blackbody reported by the COBE  FIRAS experiment  and the further refinements projected with CMB-S4 and proposed space experiments highlight a new pathway for future exploration.


 There are many papers on generating spectral distortions in the early universe: for example by decaying dark matter, evaporating primordial black holes, and even more exotic phenomenology. However it is frustrating to realize that there are no guaranteed cosmological signals for such exotica. Hence exploration of the unknown becomes difficult to justify when in competition with missions that advertise  guaranteed science goals. This is despite the fact that the historical record demonstrates that such a new horizon  approach is an invaluable part of humanity's drive to explore the universe with the prospect of vast but unforeseeable returns. In fact by being sufficiently ambitious, guaranteed science central to cosmology is achievable via a CMB spectral distortion experiment.
 
 More specifically, spectral distortions  provide a unique window that exposes the physics of the very early universe, and provide a fundamental test of the Big Bang theory, if one can overcome the various foregrounds.
Sources of spectral distortion arise in the pre-recombination and post-recombination epochs where  spectral distortions  emerge as a   combination  \citep{chluba16} of  late epoch  y-distortions \citep{1969Ap&SS...4..301Z} and early epoch
$\mu$-distortions \citep{1970Ap&SS...7...20S}   in  the standard $\Lambda$CDM model, along with a transition phase \cite{chluba}.

 The former measures energy injection in the radiation-dominated era,  after the first months, when thermalization into the cosmic blackbody radiation  is  no longer efficient. There are inevitable spectral distortions that arise  from energy injection via  damping of adiabatic baryon fluctuations if cold dark matter is the dominant form of dark matter \citep{1994ApJ...430L...5H}. 
 
 Cold  
 dark matter   generically  describes many types of dark matter, spanning  bosonic axions of mass  microelectron volts, motivated by QCD  and resolution of the strong CP problem,  to weakly interacting massive fermions (WIMPS), motivated by supersymmetry, of  mass up to 100 TeV  and yielding the observed dark matter fraction for typical weak cross-sections, and even to ultralight axions, motivated (controversially) by the prevalence of dwarf galaxy cores.

 Unitarity  arguments limit the WIMP mass: at higher masses,  the cross-section is found to be too low and WIMPs are overproduced. However refined estimates allow a window at the few TeV mass scale where massive particles are  well-motivated candidates for dark matter that have not so far been detected despite intensive searches \cite{2019PhRvD.100d3029S}.
 Verifying  the $\Lambda$CDM model is therefore of the highest importance. Given the vagaries of dark matter direct or indirect detection searches,
 measurement of the early universe CMB spectral distortions at the dwarf galaxy precursor scale may provide a unique pathway towards this goal.

	\begin{figure}[!hbt]
		\centering 
		\begin{center}
	\includegraphics[width=0.4\textwidth,
]
{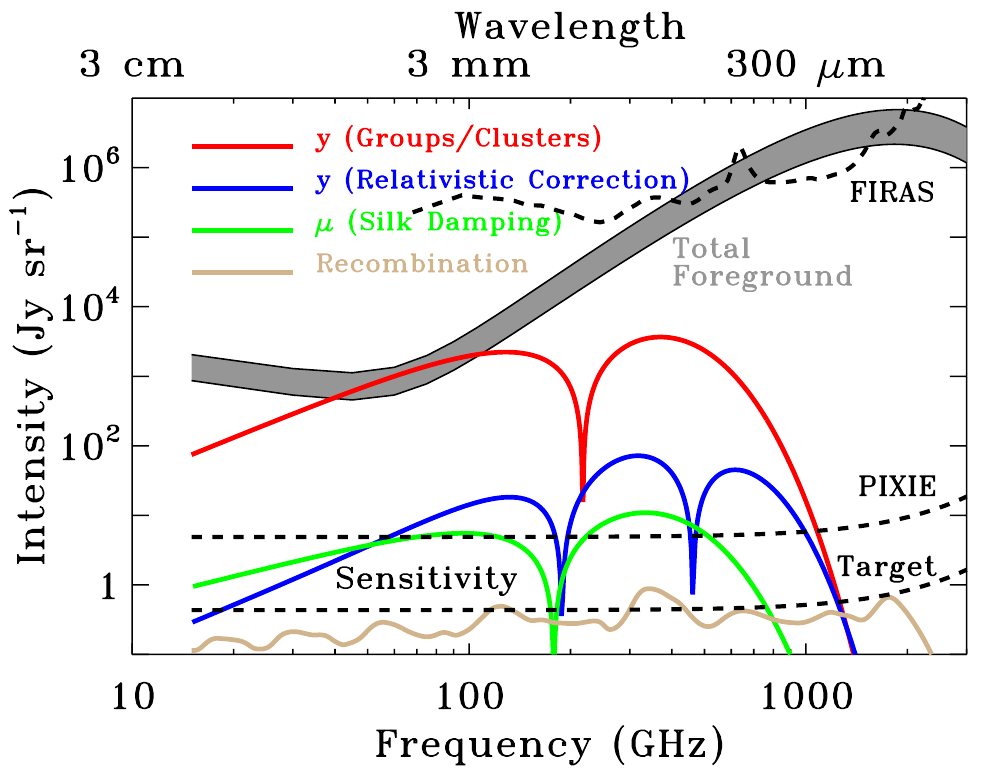}		
\caption{
Spectral distortions versus frequency, demonstrating the infrared foregrounds, the standard and relativistic y and $\mu$ spectral distortions and the hydrogen and helium recombination lines.
Also shown is the proposed PIXIE sensitivity and the target sensitivity needed for a guaranteed minimal science return. 
Adapted from \cite{desjacques}.
}
\end{center}
\label{SD} 
\end{figure}
 Dwarf galaxies are a key frontier for dark matter cosmology. New generations of telescopes and surveys  allow their detection at unprecedented surface brightnesses and distances.  They provide unique laboratories for harboring dark matter largely uncontaminated by the presence of stars. Yet their numbers are largely unknown, presenting an important diagnostic for cosmology, one that indeed has motivated many speculations about the role of astroparticle physics in controlling their abundance and other properties. 
 
 More information on their abundance, especially in terms of the primordial power spectrum, would be invaluable.  This issue can be resolved via $\mu$-type spectral distortions.  One can probe the abundance of dwarf galaxy precursor fluctuations by probing the primordial fluctuation spectrum on scales that are only accessible to spectral distortions. Direct  measurements of CMB temperature fluctuations are damped away on these scales. But the damping comes with a price: energy generation, visible uniquely via $\mu$ spectral distortions.

However because only one number is measured  for energy input in the $\mu$ era, and one cannot distinguish between the various competing and hypothetical energy inputs.  The only guaranteed return is that of testing the CDM prediction of  adiabatic fluctuation damping, which  generates dissipation on dwarf galaxy scales  in the radiation-dominated  era, after the epoch of thermalization at $\sim 10^6 $sec or $z\sim 10^7$ \cite{zaldarriaga}.  Detection of the resulting $\mu$-distortions  would be a major cosmological  result. but 
requires a sensitivity of $10^{-9},$  some four orders of magnitude improvement on FIRAS.

Detection of the damping of dwarf galaxy fluctuations would allow a unique glimpse of their cumulative power and test one of the most important predictions of the cold dark matter hypothesis, in precisely the mass range available to  $\mu$--type spectral distortions. 
These  probe  wave numbers spanning
$50$ to $500 \rm \, Mpc^{-1 }$ corresponding to mass scales $10^7$ to $10^4 \rm  M_\odot $  and generated via the baryon damping power  at  $10^4 \ltsim z\ltsim 10^7,$
on otherwise unobservable dwarf galaxy scales.  Baryon damping has been measured via the CMB acoustic oscillations on galaxy cluster scales, and it would be a fundamental  extension  and confirmation of the standard cosmological model to detect such effects emerging down to dwarf  galaxy scales. This is important for distinguishing CDM from such rivals as \cite{profumoDM} WDM (warm dark matter), SIDM (self-interacting dark matter), FDM (fuzzy  or ultralight dark matter) and other models with power spectrum cut-offs.

%

However one can do even more. Going one order of magnitude further in sensitivity 
\cite{2020JCAP...01..050S}
would enable detection of hydrogen and helium recombination lines at $z \sim 10^3-10^4 $,
including the Paschen,  Balmer and Lyman lines,
 and provide the deepest images ever attainable of the emerging universe.  Detection of the recombination lines in the energy spectrum of the cosmic microwave background would be  a clean probe of the first  atoms formed in the Universe and  a "holy grail" for understanding  the physics of the early Universe. The imprints of the recombination lines cause tiny deviations in the blackbody spectrum of the CMB with highly predictable features  determined by  atomic physics and the cosmic constituents of the Universe (Fig.2).

  The (re)combination spectral lines of hydrogen provide unique spectral distortions  produced 380,000 yrs after the Big Bang. We currently measure atomic hydrogen directly  in absorption against quasars to $z \sim 7.5$ \cite{banados}.
 Most recently, we may have controversially detected  redshifted  21 cm absorption of H atoms  to $ z\sim 17 $ in the EDGES experiment. 
  Measuring hydrogen atoms at $ z\sim 1000$, long before the first stars formed, would be an unprecedented advance in our confirming the physics of the Big Bang.

The most remarkable goal would be the first direct measure of the primordial helium abundance $Y_p$.  Planck does this indirectly, but $Y_p$ is degenerate with the number of relativistic species at BBN at roughly the 20\% level.
The  (re)combination spectral lines of helium take us back to an epoch long before any helium-synthesizing stars had formed, thereby bypassing the major uncertainty in direct measurements of $Y_p$.  We can  measure helium recombination lines back to  $z ~\sim 2000$ for $He^+ $ and 6000 for $He^{++}.$ Detection  is inevitable with enough sensitivity and would confirm the production of helium long before any stars had formed, and probe the first atoms in the universe some  20,000 to 100,000  years after the Big Bang.

 One needs a far infrared space telescope with an interferometer dedicated to spectral distortions and a large improvement on
PIXIE sensitivity, of order 10-100. The proposed but unfunded PIXIE is a dual input Martin-Puplett interferometer with two entrances, one looking at a  reference blackbody and 
 one focal plane  detector looking at  the sky with the  mirror of a  single
55 cm diameter telescope. PIXIE  was  rejected twice by the NASA MIDEX program, in  part because it lacked the sensitivity  for a definitive detection of CMB spectral distortions at the predicted  level of $\mu\sim 10^{-9}.$ 
  
 The advantage of a lunar platform in a dark crater would allow  construction of  an enhanced PIXIE, to  enable imaging  on say  degree angular scales at the desired sensitivity by increasing the number of focal plane detectors   and with a larger telescope, reduced bandwidth and an external reference blackbody. All of these design enhancements would at least guarantee the 
 $\mu$-distortion detection \cite{maillard} and open the way to achieving even more sensitive measurements.

\section{Dark matter}
The lack of success in identifying a particle dark matter candidate generated in  beyond-the-standard-model scenarios has motivated exploration of dark matter   that builds on known physics in the form of primordial black holes. The arguments for such a form of dark matter were boosted by the early LIGO results of anomalous numbers of massive black hole candidates, although it is now  generally accepted that  astrophysical mechanisms can account for current LIGO detections and that theoretical estimates of 
primordial black hole mergers in the mass range 1-100 M$_\odot$ can most plausibly account for 
$f_{PBH}$
of order 1\% of the dark matter \cite{2017PhRvD..96l3523A}. All of the dark matter in stellar mass PBHs can even be accommodated according to one recent estimate of the predicted LIGO/Virgo merger rate \cite{2020arXiv200703565J}.
 
 Recent gravitational wave events in the mass gaps where canonical black holes cannot form (2-5 M$_\odot$ and 
 50-120 M$_\odot$ may controversially lend credence to PBH mergers, although astrophysical mechanisms are also possible for rare events.

The strongest observational constraints  come  from microlensing searches, which actually constrain $f_{PBH}\ltsim 0.2$ \cite{carr} below  (approximately) $\sim 10\rm M_\odot.$
Complementary, although less stringent,  dynamical  limits from ultradiffuse galaxies constrain PBHs in  the range $1-100 \rm M_\odot$ from providing all of the dark matter \cite{stegmann}. 
There are important  implications for dwarf galaxies, the latest 
high resolution simulations  allowing $f_{PBH}\sim 0.01$ and  $M_{pbh}\sim 25-1000 \
\rm M_\odot$ to 
simultaneously account for both ultradiffuse dark matter profile diversity and dark matter cores  in faint dwarfs \cite{boldrini}.

 As for more massive PBHs, accretion is argued to  result in strong constraints from CMB temperature and polarization signatures \cite{ali}. The resulting limits eliminate  any significant contribution  to the DM density by PBHs  in the range 
$M_{BH}\gtsim 100\rm M_\odot$, 
and  even  stronger  limits are claimed \cite{serpico}.
Accretion onto the PBHs during the epoch of matter domination also results in heating of the IGM, and generates Compton y-distortions of the CMB spectrum that are currently unobservable.  

However Bondi-Hoyle-Lyttleton  accretion rates are highly uncertain especially  in the generic  limits of inhomogeneous and anisotropic  accretion,  by some two orders of magnitude once realistic geometries and inhomogeneities \cite{waters} 
 are included.
Indeed, accretion 
may even be dominated by  mechanical feedback from BH-driven outflows
\cite{2019MNRAS.482.4642Z, 2020arXiv200411224B}.

 Rather, attention has shifted to the extended mass window from $10^{-17}$ to $10^{-11}$ M$_\odot$ where asteroid-mass PBHs remain a viable dark matter option.
 Extended  limits  are obtained via gravitational microlensing down  to lens sizes of the order of the  Einstein radius  most notably in M31, where the lens are typically giant stars, corresponding to PBH masses   $\sim 10^{-10}\rm M_\odot$ \cite{smyth},
 \cite{hirataA}. 

  
  Hawking evaporation limits from the diffuse gamma ray background  and especially  high energy positron  fluxes limit the PBH mass to above $10^{-17}\rm gm$  \cite{boudaud}.
  Uncertainty about the quantum gravity aspects of smaller black holes means that under certain conditions, such as extremal spin or high charge, Hawking evaporation may be avoided
\cite{pacheco}. Moreover only an infinitesimal  fraction $(1+z_{eq})/(1+z_{pl})$ need be stable and long-lived, where $z_{eq}$ and $z_{pl}$  are the redshifts of matter-radiation equality and the Planck epoch, respectively. This means that even (extremal) Planck mass PBHs provide a possible dark matter solution, consistent with all known constraints  \cite{profumo}. 
Production requires tuning of initial conditions, but at their production epoch, PBHs are exceedingly rare because the universe is highly radiation-dominated.

Primordial black holes are the only dark matter candidate that avoids introducing new physics. One  needs  boosted scale-dependent  and possibly non-gaussian initial conditions, available in many inflationary models,  leading to spectral power boosts in the primordial density fluctuation distribution.

 Particle dark matter has  the likely possibility of annihilation or decay as a potential detection signal. The typically high energy products include gamma rays as well as high energy neutrinos, positrons and antiprotons.
Identification of a weakly interacting  dark matter particle candidate  is the holy grail of particle astrophysics. However searches in our galaxy and of nearby dwarf galaxies and galaxy clusters, all known dark matter repositories, have had little success.
 
 The dark ages offer a unique environment for exploring long-lived particle annihilations or decays, the 21cm signal  fluctuations being  sensitive to electromagnetic energy injection via precision spin temperature measurements of intergalactic hydrogen clouds at very high redshift. 
 However the low frequency terrestrial foregrounds are a severe problem, in the absence of space or lunar projects.  
 
 There has been one claim of a global signature due to 
  21-cm hyperfine absorption of cosmic microwave background photons at $z\sim 17$ or at 78 MHz
  \cite{edges}. If of cosmological origin,  the HI spin temperature  must be coupled to the kinetic temperature of the ambient hydrogen at this redshift via scattering of Lyman alpha  photons emitted by the first massive stars, but the magnitude of the signal is about twice as  large as  predicted by existing cosmological models \cite{barkana}. 
 The result has been criticized and  awaits independent confirmation \cite{2018Natur.564E..32H}. 
 
  In addition, there is  no satisfactory theoretical explanation of the magnitude and line profile of the reported absorption feature, effectively requiring extra cooling of the baryons,  either in standard cosmology 
  or via scattering of dark matter particles with baryons \cite{2019PhRvD.100b3528C}.
 Despite these uncertainties, the sensitivity of the dark ages to the most modest energy injection  provides potentially  large  improvements over the best current limits on exotic physics, in a regime that is likely to be uncontaminated by  astrophysical processes.


Another early signal that precedes structure formation arises from the abrupt decline in sound speed of the baryonic matter at last scattering, from  $\Delta_\zeta [(1+z_{LSS})/(1+z_{eq})]^{1/2} c/\sqrt{3}\sim 30 \rm \, km s^{-1}$ in the fluctuations  to 
$\sim 3\,$km/s in the ambient baryonic fluid.
Here $\Delta_\zeta (k)^2=2.4 \times 10^{-9}$ is the  primordial curvature ($\zeta$) fluctuation variance per ln$k$.
Consequently 
baryonic  matter fluctuations have a highly supersonic velocity dispersion  after recombination. This rapidly dissipates as fluctuations collapse but partially suppresses the growth of matter fluctuations on small scales, modifying      the small-scale matter power spectrum at high redshift, and  leaves an imprint on baryon acoustic oscillations and the epoch of first galaxy formation \cite{hirata}. 

Dark matter models that suppress small-scale power (including  sterile neutrino dark matter \cite{boyarsky} and  self-interacting dark matter \cite{tulin}) can modify this modulation,  allowing a possible test of  LCDM predictions and different  DM models \cite{fialkov}. These can be probed with future 21 cm observations of the universe at redshifts 10-50, including lunar projects.

 \section {Galaxy and supermassive black hole formation}
 Intermediate mass black holes   in the mass range $10^3-10^5\rm \, M_\odot$  are  possibly required for seeding supermassive black holes (SMBH)  especially at high redshift \cite{inayoshi}.
Eddington-limited accretion fails if one begins with stellar mass seeds $\ltsim 50 \rm M_\odot.$  The e-folding timescale for growth, the Salpeter time, is 
$\epsilon M_{BH}c^2/L_{Edd}=\epsilon\sigma_T c/(4\pi Gm_p), $ 
or $45.10^7\epsilon$ yr. Here, in a steady state, the Eddington luminosity  $L_{Edd}=4\pi GM_{BH}m_pc^2/\sigma_T,$ sets the maximum  black hole mass  in which the radiation pressure gradient balances gravitational force and 
 $\epsilon\sim 0.1 $ is the fraction of accreting rest mass that is converted into radiation.
 The Eddington-limited  accretion rate
 fails  to give enough growth e-foldings at high redshift, when the age of the universe is a few hundred million  years,  to account for the most massive SMBHs. For example at z=7, the age of the universe was only 700 Myr.
 Eddington-limited growth rate is too  low to permit formation of SMBH with masses as high as observed at high redshift, $\sim 10^{10} \rm M_\odot.$

 Super-Eddington accretion offers  a possible solution, involving highly anisotropic  and/or non-equilibrium  accretion, possibly visible via flaring, or the changing-look phenomenon. However the duty cycle is unknown, and  
 any observational evidence for super-Eddington accretion on AGN scales remains sparse   \cite{2020MNRAS.492.4058P}.

Mergers of IMBH  seeds provide an alternative resolution, and associated  gravitational wave signals should be detectable with LISA. 
Gas accretion onto the inferred IMBH seeds during the dark ages  would leave a unique signature in the 21cm signal. 
Such excess power potentially shows up in the 21 cm autocorrelation function, uniquely probed via very low frequency 21 cm dark ages astronomy
\cite{2018JCAP...05..017B, 2020JCAP...01E.001B}.

\begin{figure}[!hbt]
\centering
\begin{center}
\includegraphics[width=0.5\textwidth,
]
{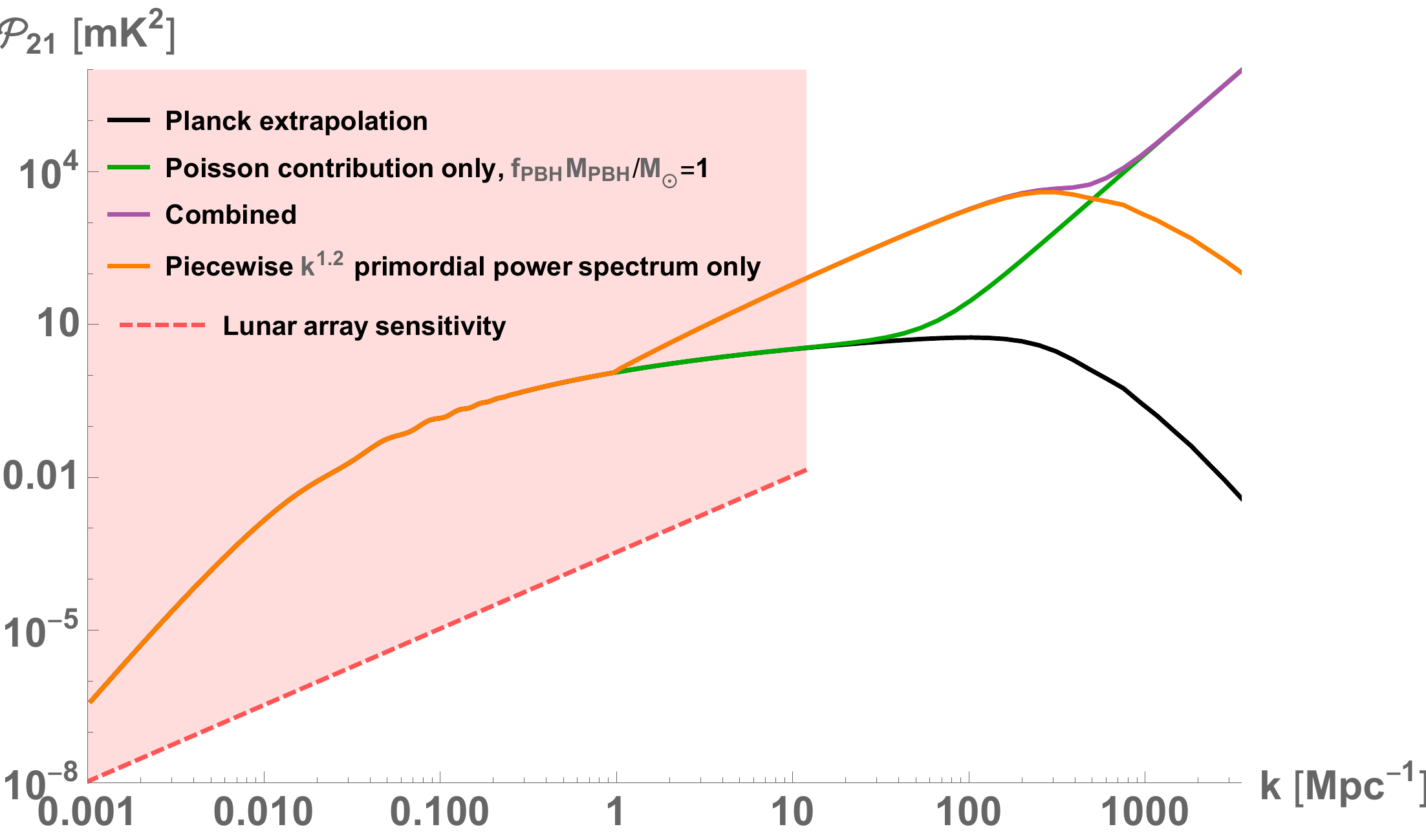}
\caption{
A rough estimate of the sensitivity of a possible configuration
for a radio interferometer on the far side of the moon is
shown by the red dashed line.
The 21cm power spectrum at redshift 50 for the scenario where 100M$_\odot$
PBHs are produced with abundance $10^{-4}$ relative to the dark matter density
The 21cm signal prediction (orange) takes into account
the boost in the primordial power spectrum, The
Poisson contribution (green), and  the combined result (purple) are shown  . In
black is the 21cm power spectrum produced by extrapolating the
primordial power spectrum measured by Planck to small scales. From \cite{cole}.
}
\label{figcole2}
\end{center}
\squeezeup
\end{figure}

 A 21cm lunar radio array could have the sensitivity to detect their signature and constrain their properties,  modifying the standard (LCDM-based) view of the high redshift universe and  complementing CMB  probes.  Poisson fluctuations are detectable in the 21cm power spectrum if the mass fraction of seed IMBH is as high as $\sim 10^{-4} $ (Fig.~\ref{figcole2}).
 
 Another  and complementary approach to massive black hole science is the following. One could exploit the frequency gap    between gravitational waves detectable by  LISA  (sensitive to IMBH mergers of masses $\gtsim \rm\, 10^5 M_\odot$)
 and LIGO/Virgo/Einstein. (sensitive to BH  masses  $\ltsim 100\rm \,M_\odot$) to probe the build-up
 of massive black holes. 
 Construction of a gravitational wave  lunar interferometer that would operate at subHz frequencies  over a baseline of around 100 km would enable the range of frequencies that cannot be easily accessed  by  terrestrial or space-based detectors \cite{2020arXiv200708550J}.

  
\section{Exoplanets}
We now turn to the exoplanet science capabilities of a very large  lunar telescope. Current hopes are pinned on LUVOIR \cite{luvoir},  a proposed 10-15m imaging space telescope.
 with capabilities of observing exoplanet spectra of $\sim$50 Earthlike planets. These targets are  defined to have rocky cores and be in habitable zones of the host stars. These latter are  nearby main sequence stars attainable over the course of a 25 year mission.
The total sample size for LUVOIR (15m)  for rocky core exoplanets in habitable zones  is taken to be 50, since it is capable of searching only a zone  25 pc distant from the sun with sufficient precision.


Now lets compare these goals to a lunar version of OWL.  OWL was a concept of a 100 m telescope \cite{gilmozzi}  that was abandoned in favour of the 39m ELT, in part because of engineering  constraints that made it unfeasible on any terrestrial site. Such constraints do not apply on the Moon, with lower gravity and no winds, and combined with the huge advantage of no atmosphere such a concept merits reconsideration. It will have  many unique applications, for example in cosmology, but 
it is in enhancing exoplanet yields that size is of paramount importance.
  
 Using the scaling with telescope diameter found in \cite{stark2014maximizing} for noise dominated by zodiacal light, $N\propto D^{1.8}$, we find that $\sim 1000$ planets could be imaged (but not resolved) by  a 100 m telescope with sufficient signal-to-noise. 
 Such a yield is a huge statistical gain over LUVOIR and brings significantly enhanced chances of finding atmospheric biosignatures for exoplanets within a search radius of 250 pc. .
 \begin{figure}[!hbt]
 
 \squeezeup
\squeezeup
\squeezeup
\squeezeup
\squeezeup
\squeezeup
\squeezeup

\squeezeup
\squeezeup
		\centering
		\includegraphics[width=0.5\textwidth]
		{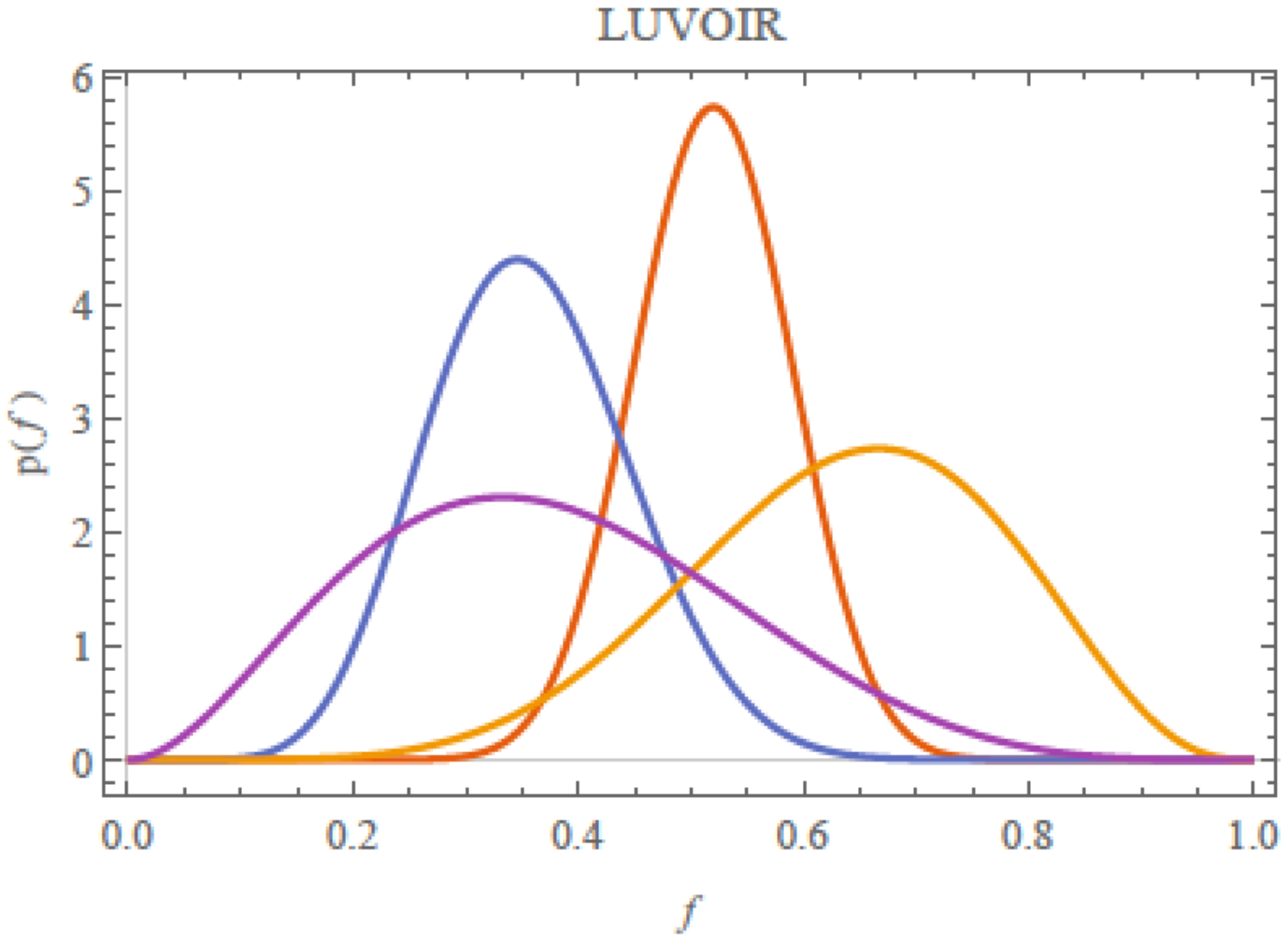} 

\squeezeup
\squeezeup
\squeezeup
\squeezeup
\squeezeup
\squeezeup
\squeezeup

\squeezeup
\squeezeup		
		\caption{Inferred values of fraction of planets which attain each successive stage of evolution, assuming optimistically high values of $f_\text{chem}=0.5$, $f_\text{photo}=0.4$, $f_\text{multi}=0.7$, and $f_\text{tech}=0.6$.  The total sample size for LUVOIR (15m)  for rocky core exoplanets in habitable zones  is taken to be 50.   Color coding is red, diagnostics of necessary chemical synthesis; blue, photosynthesis; orange, multicelluarity;  purple, technology signatures. From \cite{sandora}.}
		\label{sandora1}
	\end{figure}

 \begin{figure}[!hbt]
\squeezeup
\squeezeup
\squeezeup
\squeezeup
\squeezeup
\squeezeup
\squeezeup

\squeezeup
\squeezeup
		\centering
		\includegraphics[width=0.5\textwidth
]
{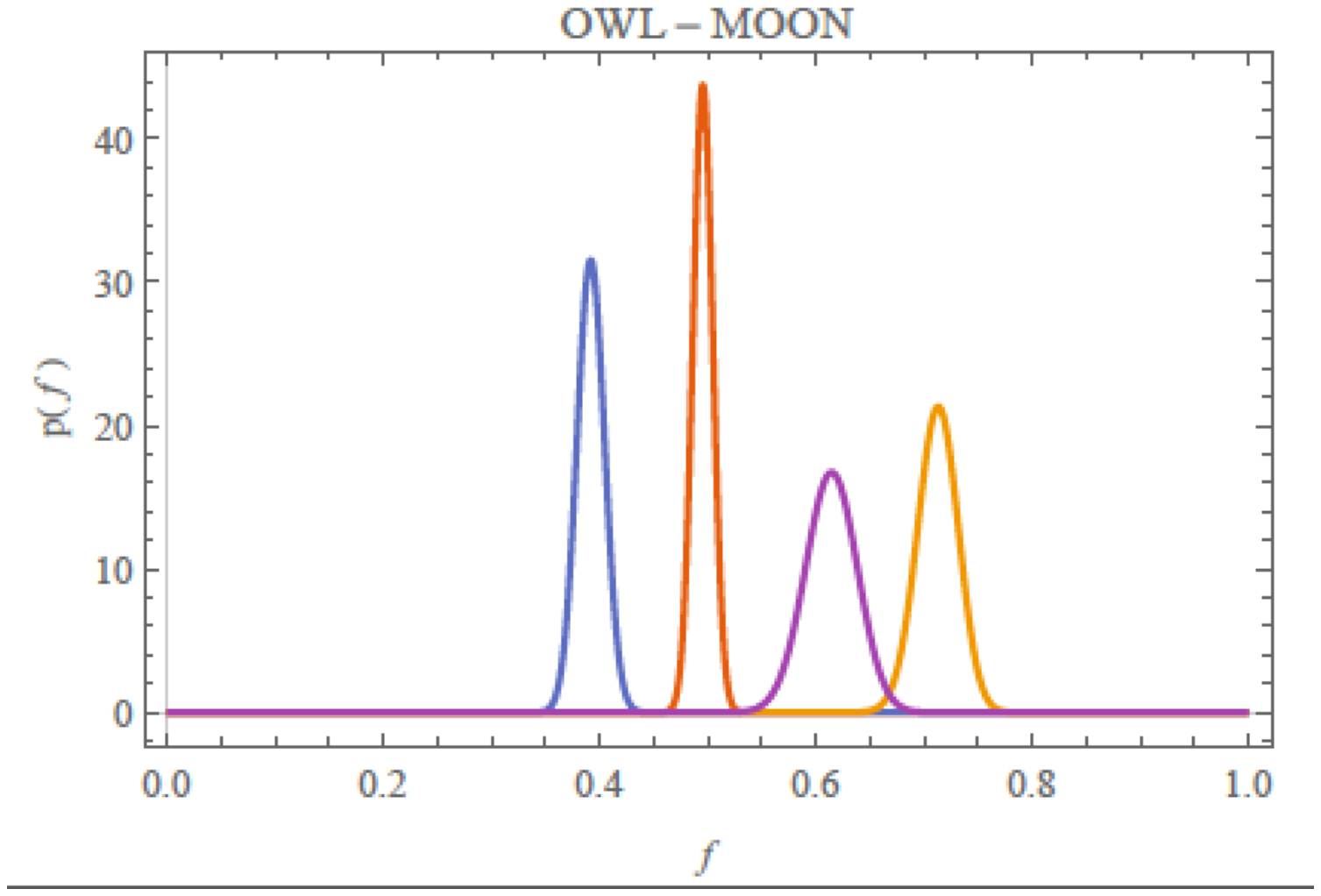}

\squeezeup
\squeezeup
\squeezeup
\squeezeup
\squeezeup
\squeezeup
\squeezeup

\squeezeup
\squeezeup
		\caption{Inferred values of fraction of planets which attain each successive stage of evolution, assuming optimistically high values of $f_\text{life}=0.5$, $f_\text{photo}=0.4$, $f_\text{multi}=0.7$, and $f_\text{tech}=0.6$.  The total sample size for OWL  (100m) is taken to be 1000. Color coding is red, diagnostics of chemosynthesis; blue, photosynthesis; orange, multicelluarity;  purple, technology signatures. From \cite{sandora}.}
		\label{sandora2}
	\end{figure}

The limiting factor, however, may be the image contrast ratio: if an orbital star shade is employed, repositioning times may be long.  As opposed to  telescopes situated on Earth, the moon's atmosphere will not introduce a contrast floor, potentially granting access to habitable zones around sun-like stars as well as M dwarfs.

Extension to NIR wavelengths 
would give  access to spectral peaks of many chemical species, including water, methane, carbon monoxide and dioxide, and oxygen.  Observed together, these have been argued to strongly indicate the presence of life, and would be feasible with 
 a 100 m telescope for 
photon-limited $t\propto d^{-2}$ noise.  
Additionally, IR excesses indicating the presence of excess planetary heat are prime candidates for technosignatures.

Another way of quantifying the yields of different telescope missions is by the certainty with which we can infer the fraction of planets that have various biosignatures.  This is typified by the sequence of planets which possess life, and  the fraction of those which have developed photosynthesis, then multicellularity, and finally technology.   
Any form of life is expected to be accompanied by chemosynthesis, the rearrangement of chemical matter
Chemosynthesis is the minimal requirement,  essentially equivalent to the presence of
gaseous byproducts  that are indicative
of life, chief among those being methane CH4, as well as water, CO2 and N2.

The
next step is photosynthesis, the harvesting of energy from the planet's host star. This
innovation provides a mechanism to harvest the dominant source of free energy on a
planet.  Molecular oxygen is a signpost, along with ozone, NH and OH. Of course, chlorophyll and the asociated red spectral edge 
are key tracers, especially via  scattering-induced polarization   off vegetation.
A following step in complexity is the  evolution of multicellularity as a key prerequisite for intelligent life. Examples might include extremophiles, plant biomass signatures via phased signatures of scattered light.
Lastly, the
detection of technological civilizations, both on the level of and beyond the level that  the Earth has achieved, is a final consideration
for the feasibility of  discovery signatures in exoplanet atmospheres.

Precise biosignatures corresponding to each of these stages of development can be estimated, but planets with each of these characteristics are expected to be successively rarer.  The extent to which we can infer the fraction that attain each biosignature level will be  limited by the sample size of the exoplanet yield.  This is displayed in Figs. \ref{sandora1} and \ref{sandora2}.

The number of exoplanets that needs to be targeted is debatable. However a field of, say, tens of targets with rocky cores in habitable zones is almost certainly  too small to have any reasonable odds for success. Ideally  
a search basis of a thousand or more such exoplanets would seem to be  required for  detailed multimessenger spectroscopic follow-up. This  only becomes available with the aperture provided by a lunar megatelescope.

 A secondary goal of such a telescope would be  detection of the first stars in the universe.These are rare but luminous, and could be visible  with enough sensitivity as Population III  star clusters \cite{2015MNRAS.449.3057Z} or  transient phenomena  such as supernovae
 \cite{2018ApJS..234...41W},  or even as  individual very massive stars \cite{2020arXiv200702946S}.

 \section{Final comments}
 There are many issues that need to be explored before a lunar telescope can be built. The effects of lunar dust and solar cosmic rays on electrical and mechanical components need to be studied. Data collection and analysis presents a challenge along with deployment of robotic capabilities and risks of human servicing. 
 
 While the lunar surface is a natural location for low frequency radio interferometry, with a stable platform,  no ionosphere, and far-side shielding from  terrestrial interference, the case for infrared and optical telescopes is less obvious. The lack of atmosphere provides a unique space-like advantage, and scaling in size offers 
 prospects for imaging and spectroscopy  at unprecedented light gathering power. Another key science goal  involves   construction of a gravitational wave  lunar interferometer that would operate at subHz frequencies  filling a gap between terrestrial and space-based detectors.  

Unique advances in cosmology  are feasible with lunar telescopes that will probe  our cosmic origins within the first thousands of years after the Big Bang, and even as far back as  the epoch of inflation.
 
 There is a down side, most notably that of charged dust particles. These could affect IR/optical/UV telescope performance. The Moon is embedded in a tenuous impact-generated dust cloud detected by the Apollo lunar horizon glow 
 \cite{2015Natur.522..324H}. Lunar regolith particles are lofted to   heights that depend on electrostatic forces,  with dust charge accumulation  varying  with lunar day/night  and solar wind/coronal flaring activity
 \cite{2007RvGeo..45.2006C}. 
 The lack of direct solar exposure may mitigate dust lofting in dark craters although micrometeorite impacts will generate some surface particle  activity.
 



 However the  budget needed for lunar telescopes, while overwhelming  for stand-alone projects, and hence financially  unacceptable, remains only a small fraction of the overall lunar exploration budget.  Very large space  telescopes  as free space observatories can be designed to be extremely competitive with the goals outlined here
 \cite{2004SpPol..20...99L}. but 
 are likely to be prohibitively expensive as stand-alone projects.
 

 Just as the Hubble Space Telescope might not have emerged without the space infrastructure developed for the International Space Station and the Space Shuttle, at a cost of order 5\% of the total budget,
 we  can be reasonably optimistic that lunar telescopes  will have a niche alongside, and as a minor component of, the
  political priorities that drive human exploration and commercial exploitation of the Moon.

\vfill
\acknowledgments
I thank  my collaborators for invaluable discussions of many of the topics covered here,  including Philippa Cole, Jean-Pierre Maillard and McCullen Sandora. 


\end{document}